\documentclass[epj]{svjour}
\usepackage{graphics}
\usepackage{paralist}
\usepackage{acronym}
\usepackage{natbib}
\setcitestyle{numbers,square,comma}
\usepackage{pstricks}
\usepackage{array}
\usepackage{tabularx}
\usepackage{graphicx}
\usepackage{color}
\usepackage{psfrag}
\usepackage{amsmath}
\usepackage{amssymb}
\usepackage{psfrag}
\usepackage{textcomp}
\usepackage{longtable}
\usepackage{multirow}
\usepackage{braket}
\usepackage[mathscr]{euscript}
\usepackage{dcolumn}
\usepackage{bm}
\usepackage{hyperref}
\hypersetup{letterpaper, colorlinks=true, urlcolor=blue, citecolor=blue, linkcolor=blue, breaklinks=true}
\usepackage{breakurl}

\newcommand{\der}[1]{\textrm{d}#1}

\newcommand{\meas}[2]{$#1~\textrm{#2}$}
\newcommand{\nuc}[2]{$^{#1}\textrm{#2}$}

\newcommand{\nimout}[1]{\ac{NIM} output~#1}
\newcommand{\nimin}[1] {\ac{NIM} input~#1}
\newcommand{\eclin}[1] {\ac{ECL} input~#1}

\newcommand{\nimins}[2] {\ac{NIM} inputs~#1--#2}
\newcommand{\eclins}[2] {\ac{ECL} inputs~#1--#2}

\newcommand{\rxnfull}[6]{$^{#1}\textrm{#2} \left( #3 , #4 \right) ^{#5}\textrm{#6}$}

\newcommand{\ee}[2]{#1 \times 10^{#2}}
\newcommand{\wg}{\omega \gamma}

\newcommand{\figref}[1]{Fig.~\ref{#1}}
\newcommand{\eqnref}[1]{Eqn.~\eqref{#1}}
\newcommand{\tableref}[1]{Table~\ref{#1}}
\newcommand{\secref}[1]{Sect.~\ref{#1}}
\newcommand{\citeref}[1]{Ref.~\cite{#1}}

\newcommand{\nlost}{n_{\text{lost}}}
\newcommand{\of}[1]{\left( #1 \right)}

\newcommand{\panel}[1]{Panel~(#1)}

\newcommand{\ecm}{E_{\text{c}.\text{m}.}}

\graphicspath{{./figures/}}

\begin{document}
\title{Design and commissioning of a timestamp-based data acquisition system for the DRAGON recoil mass separator}
\authorrunning{G. Christian \emph{et al.}}
\titlerunning {Design and commissioning of a timestamp-based data acquisition system for DRAGON}

\author{
		G. Christian\inst{1}  \thanks{gchristian@triumf.ca}  \and 
		C. Akers\inst{1,2}    \and
		D. Connolly\inst{3}   \and
		J. Fallis\inst{1}     \and
		D. Hutcheon\inst{1}   \and
		K. Olchanski\inst{1}  \and
		C. Ruiz\inst{1}
} 


%
%

\institute{{\sc triumf}, 4004 Wesbrook Mall, Vancouver, BC V6T 2A3, Canada \and Department of Physics, University of York, Heslington, York YO10 5DD, UK \and Department of Physics, Colorado School of Mines, 1523 Illinois Street, Golden, CO 80401, USA } 
\date{Received: / Revised version: }
%
\abstract{
	The DRAGON recoil mass separator at {\sc Triumf} exists to study radiative proton and alpha capture reactions, which are important in a variety of astrophysical scenarios. DRAGON experiments require a data acquisition system that can be triggered on either reaction product ($\gamma$ ray or heavy ion), with the additional requirement of being able to promptly recognize coincidence events in an online environment. To this end, we have designed and implemented a new data acquisition system for DRAGON which consists of two independently triggered readouts. Events from both systems are recorded with timestamps from a $20~${}MHz clock that are used to tag coincidences in the earliest possible stage of the data analysis. Here we report on the design, implementation, and commissioning of the new DRAGON data acquisition system, including the hardware, trigger logic, coincidence reconstruction algorithm, and live time considerations. We also discuss the results of an experiment commissioning the new system, which measured the strength of the $\ecm = 1113$ keV resonance in the $^{20}$Ne$\left( p , \gamma \right)^{21}$Na radiative proton capture reaction.
} 
\maketitle

\newcommand{\allcaps}[1]{\MakeUppercase{#1}}

\newcommand{\code}[1]{\texttt{#1}}
\newcommand{\ROOT}{{\sc Root}}
\newcommand{\GEANT}[1]{{\sc Geant#1}}
\newcommand{\cpp}{C{}\texttt{++}}
\newcommand{\triumf}{{\sc Triumf}}
\newcommand{\isac}{\allcaps{isac}-\allcaps{i}}
\newcommand{\iothirtytwo}{\allcaps{io32}}
\newcommand{\caen}{{\sc Caen}}
\newcommand{\caenv}[1]{V$#1$}
\newcommand{\logical}[1]{{\sc #1}}

\acrodef{VME}  [\allcaps{vme}]{V\allcaps{ersa}module Eurocard}
\acrodef{CAMAC}[\allcaps{camac}]{computer automated measurement and control}
\acrodef{NIM}  [\allcaps{nim}]{nuclear instrumentation module}
\acrodef{ECL}  [\allcaps{ecl}]{emitter coupled logic}
\acrodef{FPGA} [\allcaps{fpga}]{field-programmable gate array}
\acrodef{DAQ}  [\allcaps{daq}]{data acquisition}
\acrodef{DRAGON}[\allcaps{dragon}]{Detector of Recoils and Gammas of Nuclear Reactions}
\acrodef{DSSSD}[\allcaps{dsssd}]{double-sided silicon strip detector}
\acrodef{MCP}  [\allcaps{mcp}]{microchannel plate}
\acrodef{IC}   [\allcaps{ic}]{ionization chamber}
\acrodef{TOF}  [\allcaps{tof}]{time of flight}
\acrodef{BGO}  [\allcaps{bgo}]{bismuth germanate}
\acrodef{TDC}  [\allcaps{tdc}]{time to digital converter}
\acrodef{QDC}  [\allcaps{qdc}]{charge to digital converter}
\acrodef{ADC}  [\allcaps{adc}]{amplitude to digital converter}
\acrodef{CFD}  [\allcaps{cfd}]{constant fraction discriminator}
\acrodef{FIFO} [\allcaps{fifo}]{first in, first out}
\acrodef{TSC}  [\allcaps{tsc}]{timestamp counter}
\acrodef{IIS}  [\allcaps{iis}]{ion-implanted silicon}
\acrodef{NaI}  [\allcaps{n}a\allcaps{i}]{sodium iodide}
\acrodef{HPGe} [\allcaps{hpg}{e}]{high purity germanium}
\acrodef{MIDAS}[\allcaps{midas}]{Maximum Integrated Data Acquisition System}
\acrodef{RF}   [\allcaps{rf}]{radio frequency quadrupole}

\section{Introduction}
\label{sec:intro}

\subsection{The DRAGON Facility}
\label{subsec:dragon}

Radiative capture reactions typically involve the absorption of a light nucleus (typically a proton or an $\alpha$ particle) by a heavy one, followed by $\gamma$-ray emission.  These reactions are important in a variety of astrophysical scenarios such as novae~\cite{PhysRevLett.110.262502, PhysRevLett.105.152501, PhysRevC.81.045808, PhysRevLett.96.252501, PhysRevLett.90.162501, PhysRevC.88.045801}, supernovae~\cite{PhysRevC.76.035801}, X-ray bursts~\cite{6208cab5983b4072a365e4e3c7079eed, 0004-637X-735-1-40}, and quiescent stellar burning~\cite{PhysRevLett.97.242503, Schurmann2011557}. They are often difficult to study directly in the laboratory. The cross sections are low, typically on the order of picobarns to millibarns, since the relevant energies are below the Coulomb barrier. Additionally, many interesting reactions involve short-lived nuclei and can only be studied using low-intensity radioactive beams.

\begin{figure}
\centering
\includegraphics{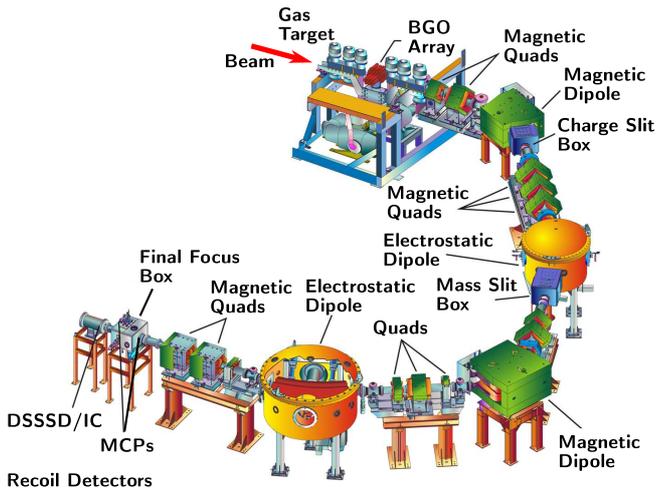}
\caption{The \ac{DRAGON} facility at \triumf{}.}
\label{fig:dragon}       
\end{figure}

The \ac{DRAGON} facility at \triumf{}~\cite{Hutcheon2003190}, shown in \figref{fig:dragon}, is a recoil mass separator that was built to study radiative capture reactions using stable and radioactive beams from the \isac{}~\cite{Laxdal2003400} facility. \ac{DRAGON} experiments are typically performed in inverse kinematics with a beam of the heavy nucleus impinging on a windowless gas target containing the lighter one.  Beam energies range from $E/A = 0.15$--\meas{1.5}{MeV}.  The products of radiative capture (recoils) are transmitted through \ac{DRAGON} and detected in a series of charged particle detectors, while unreacted beam and other products are deposited at various points along the separator's flight path.  The recoil detectors consist of a pair of \acp{MCP} to measure local \ac{TOF}~\cite{Vockenhuber2009372} and either a \ac{DSSSD}~\cite{Wrede2003619} or an \ac{IC} to measure energy loss. The $\gamma$ rays resulting from radiative capture are detected in an array of $30$ \ac{BGO} detectors surrounding the target.

For beam normalization, the target chamber houses two \ac{IIS} detectors to record elastically scattered target nuclei. In a typical experiment, the scattering rates measured in the \ac{IIS} detectors are normalized to hourly Faraday cup readings of the absolute beam current. Experiments using low-intensity and possibly unpure radioactive beams may also include a pair of \ac{NaI} scintillators, a \ac{HPGe} detector, or both. These auxiliary detectors are located near the first mass-dispersed focus, and they detect the $\gamma$ rays resulting from the decay of radioactive beam deposited onto the nearby slits. This allows a continuous determination of the beam rate and composition throughout the experiment.

In many experiments, unreacted, scattered, or charge-changed beam particles (``leaky beam'') are transmitted to the end of \ac{DRAGON} along with the recoils of interest. The rates vary depending on experimental conditions but can potentially be as much as a few thousand times the recoil rate~\cite{EPJNe20}. Hence, it is crucial that leaky beam be separable from recoils in the data analysis. In some cases, separation is possible using the signals from recoil detectors alone. In others, it is necessary to require (delayed) coincidences between the heavy ion and a $\gamma$ ray measured in the \ac{BGO} detectors.  In such experiments, a measurement of the \ac{TOF} between the $\gamma$ ray and the heavy ion (``separator \ac{TOF}'') is useful for distinguishing genuine coincidences from random background.

%

\subsection{Data Acquisition Requirements}
\label{subsec:daq}

As mentioned, identification of coincidences between the ``head'' ($\gamma$-ray) and ``tail'' (heavy-ion) detectors is important for many \ac{DRAGON} experiments. As a result, the original \ac{DRAGON} \ac{DAQ} was designed to trigger on singles events from either detector system while also identifying coincidences from hardware gating. The resulting trigger logic was rather complicated and required a moderate amount of hardware reconfiguration when changing the detector setup (for example, swapping the \ac{DSSSD} and \ac{IC}). With this system, the potential for logic problems due to human error or faulty modules was relatively high, resulting in the possibility of wasted beam time or otherwise non-optimal data sets.

In order to alleviate the problems associated with the existing coincidence logic, we have designed and implemented a new \ac{DAQ} system for \ac{DRAGON} that identifies coincidences from timestamps instead of hardware gating. In the course of doing this, we have also upgraded the digital readout from a \ac{CAMAC} system to \ac{VME} and migrated part of the trigger logic from \ac{NIM} hardware to a \ac{FPGA}. In the new set\-up, the head and tail systems are triggered and read out completely independent of each other, and coincidences are identified in the analysis stage from timestamp matching.

In this paper, we provide an overview of the new DRA\-GON \ac{DAQ} system and data analysis codes. We also discuss the results of the \ac{DAQ} commissioning experiment, which consisted of a measurement of the $\ecm = 1113$ keV resonance strength in the \rxnfull{20}{Ne}{p}{\gamma}{21}{Na} radiative proton capture reaction.

\section{Trigger Logic}
\label{subsection:triggerlogic}

\begin{figure}
\centering
\includegraphics{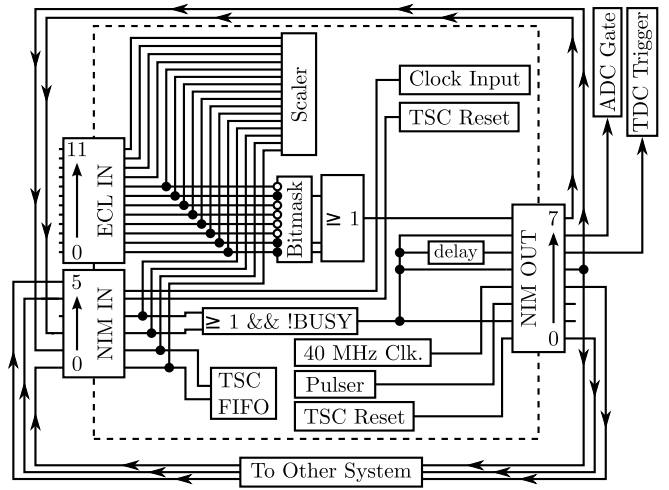}
\caption{Diagram of the generic \iothirtytwo{} \ac{FPGA} logic. See the text for further explanation.}
\label{fig:io32}
\end{figure}

The majority of the \ac{DRAGON} trigger logic and timestamping functionality is implemented in \ac{FPGA} firmware.  For this we use an \iothirtytwo{}, a general purpose \ac{VME} board designed and manufactured at \triumf{}~\cite{NIMIO32}. The \iothirtytwo{} houses an Altera Cyclone-I \ac{FPGA}~\cite{AlteraCyclone} and has input-output capabilities via sixteen \ac{NIM} and sixteen \ac{ECL} input channels and sixteen \ac{NIM} outputs. It also houses a \meas{20}{MHz} quartz oscillator crystal with an accuracy rating of $20$ parts per million.

The \ac{FPGA} logic is designed in a generic way, allowing identical firmware to be used for both the head and tail systems. A diagram of the \ac{FPGA} logic is shown is \figref{fig:io32}. \eclins{0}{11} accept trigger signals from various detectors and are all routed into a firmware scaler for rate counting, as are \nimins{0}{3}. \eclins{0}{8} are also sent through a programmable bitmask, and the \logical{or} of the unmasked channels is sent to \nimout{7} which is then routed into either \nimin{2} or \nimin{3}.  The \logical{or} of \nimin{2} and \nimin{3} is combined with an internal \logical{not} \logical{busy} condition to generate a system trigger. This causes a logic pulse to be emitted from \nimout{4}, and this is then routed back into \nimin{1} which tells the system to begin acquiring data.  The input of \nimin{1} is also sent to a \ac{FIFO} data structure that stores the \ac{TSC} value denoting when the signal arrived. These data are used for coincidence matching in the analysis stage, as  explained in \secref{sec:coincmatch}. The signal from \nimout{4} is also sent to the other system (head to tail or vice-versa). There it is connected to \nimin{0} which is also routed into the \ac{TSC} \ac{FIFO}.

A system trigger also results in signals being sent from \ac{NIM} outputs $1,$ $5,$ and $6.$  \nimout{1} emits a ``busy'' signal that remains true until cleared by a \ac{VME} register setting. This signal is not necessary to run the system, but it is often useful for debugging purposes.  \nimout{5} emits a logic pulse after a programmable time delay. This pulse is sent to the system's \ac{TDC} to act as the stop signal\footnote{
	In reality, it is not a ``stop'' that is sent to the TDC, but rather a ``trigger'' signal that must come after all of the measurements in the corresponding event. See \citeref{CAENv1190} for more details.
}.
\nimout{6} emits a pulse of programmable width which is used to gate the system's \ac{ADC} or \ac{QDC}. The \iothirtytwo{} firmware also includes a programmable pulse generator. This is a square wave of programmable frequency emitted from \nimout{2}.

The \ac{TSC} is run off a clock with \meas{20}{MHz} nominal frequency. Its size is $38$ bits, allowing \meas{\sim 3.8}{hours} of run time before it rolls over. The acquisition software also keeps track of any roll over in the $38$-bit counter, allowing the system to run indefinitely.
The exact clock frequency is set either by the quartz crystal housed on the \iothirtytwo{} board or by a signal with two times the desired clock frequency (i.e., a nominal frequency of $40$ MHz) sent into \nimin{5}. The \ac{TSC} value can be reset to zero either by writing to a \ac{VME} register or by sending a pulse to \nimin{4}. Zeroing by the \ac{VME} method causes a signal to simultaneously be emitted from \nimout{0}. Multiple boards can be run in master-slave configuration where the master is clocked off its local quartz oscillator and the slave(s) are clocked off the $40$ MHz output of the master. In this setup, the master clock is zeroed by \ac{VME} and the slave(s) by a pulse sent from \nimout{0} of the master. The result is a frequency synchronization and zero-point matching which differs only by the transit time of the zero-reset pulse (which is typically negligible). The \ac{DRAGON} system is run in such a master-slave configuration, with the head \iothirtytwo{} arbitrarily designated the master and the tail the slave.

\begin{figure}
\centering
\includegraphics{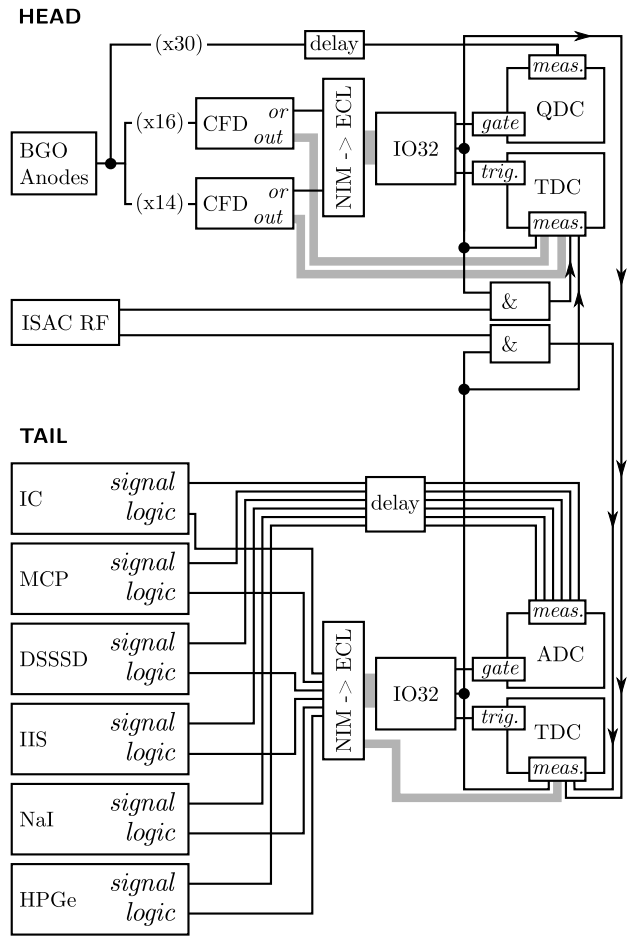}
\caption{Diagram of the timestamp-based \ac{DRAGON} trigger logic. See the text for more explanation.}
\label{fig:triggerlogic}       
\end{figure}

A diagram of the specific trigger logic used in the \ac{DRAGON} head and tail systems is shown in \figref{fig:triggerlogic}. On the head side, the anode signals from the $30$ \ac{BGO} detectors are split into analog and logic branches. The analog branch is sent through a physical time delay before going to the input of a \caen{} \caenv{792} \ac{QDC}~\cite{CAENv792} . The length of the time delay is set such that the signals arrive at the \ac{QDC} more than \meas{15}{ns} after the leading edge of the gate pulse, as required by the \ac{QDC} specifications. Signals in the logic branch are sent to a pair of \caen{} \caenv{812} \acp{CFD}~\cite{CAENv812}. The channel-by-channel \ac{CFD} outputs are sent to the inputs of a \caen{} \caenv{1190} \ac{TDC}~\cite{CAENv1190}, and the \logical{or} outputs are sent to \eclin{0} and \eclin{1} to generate the system trigger and associated signals (\ac{QDC} gate and \ac{TDC} trigger). A copy of the system trigger is sent to a measurement channel of both the head and tail \acp{TDC}, to facilitate a measurement of separator \ac{TOF}.

On the tail side, the trigger is essentially an \logical{or} of each of the heavy-ion detectors mentioned in \secref{subsec:dragon}. The outputs of each heavy-ion detector are sent through some combination of amplifiers, shapers, and discriminators (whose exact configuration varies and is outside the scope of this paper) until there is an analog signal suitable for amplitude measurement and a digital signal suitable for triggering and timing. 
The analog signals are sent to the inputs of a \caen{} \caenv{785} \ac{ADC}~\cite{CAENv785}, possibly after a physical time delay to place them within the \ac{ADC} gate. The logic signals are sent to measurement channels of a \caen{} \caenv{1190} \ac{TDC} and to \eclins{0}{7} to create the system trigger, \ac{ADC} gate, and \ac{TDC} trigger signals. As with the head system, a copy of the system trigger is sent to measurement channels of both the head and tail \acp{TDC}, resulting in a redundant measurement of separator \ac{TOF}.  In some cases, logic signals from detectors that measure incoming beam rates or composition (\ac{IIS}, \ac{NaI}, and \ac{HPGe}) may be downscaled to reduce the total trigger rate and, correspondingly, the dead time.

In both systems, a copy of the $11.8~${}MHz \isac{} \ac{RF} accelerator signal is sent to the \ac{TDC} to be used as an additional timing reference. To avoid swamping the TDC buffers with \ac{RF} pulses, the signal is gated by an adjustable-width copy of the system trigger. Typically, the gate width is set large enough that three full RF pulses are captured for every event.

\section{Data Acquisition and Analysis}
\label{subsection:analysis}

The data acquisition and online data analysis codes are both implemented as part of the \ac{MIDAS} framework~\cite{ritt1997midas}. The acquisition code is implemented in \cpp{} and employs device driver codes that are widely used at \triumf{}~\cite{daqplone}. \ac{MIDAS} transition handler priorities are used to specify the order of head and tail initialization routines at the beginning of each data-taking run. This ensures that operations which are required for timestamp matching, such as \ac{TSC} zeroing, are performed in the necessary order.

The analysis codes are also written in \cpp{} and are designed such that they can be used for both online and offline analysis using the \ROOT{} data analysis framework~\cite{Brun199781}. Each individual detector in the system is represented in a \cpp{} class, with data fields corresponding to the available measurement parameters. The various detectors in the head and tail systems are then composed into a larger class. For coincidence events, the head and tail classes are further combined as members of a single coincidence class. Such a design facilitates easy integration into the \ROOT{} framework, with the class hierarchy naturally transforming to branches and sub-branches in a \ROOT{} tree. The entire analysis suite, including the complete development history, is hosted in an online repository that is publicly viewable \cite{dragonGithub}.


\subsection{Coincidence Matching}
\label{sec:coincmatch}

\begin{figure}
\centering
\includegraphics{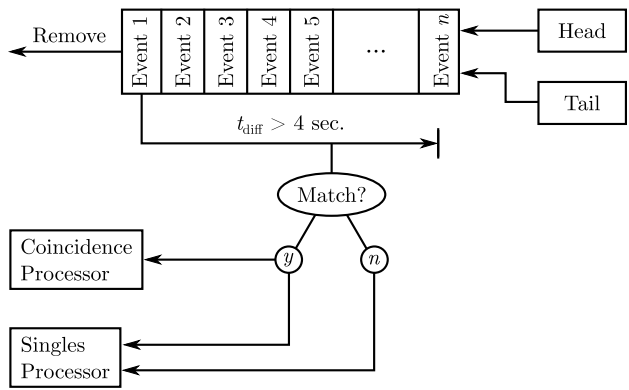}
\caption{Diagram of the coincidence matching algorithm. See the text for more information.}
\label{fig:tsmatch}       
\end{figure}

With the shift in coincidence tagging from the hardware to the analysis phase of the experiment, it was necessary to develop an algorithm that is capable of accurately identifying coincidence events in both an online and an offline environment, for all possible trigger rates. The particulars of the MIDAS system create some challenges for online identification of coincidences.  In MIDAS, event data are transferred from the ``frontend'' \ac{VME} processor to the ``backend'' analysis computer via a gigabit ethernet connection.  For efficiency reasons, transfers are made only once per second, with events buffered locally in the \ac{VME} processor in between.  As a result, events from the head and the tail frontends arrive at the backend asynchro\-nously. This is because the time of arrival is dictated by when the frontends are ready to send a packet of events, not the actual trigger time of any given event. Thus the coincidence matching algorithm must ensure that events with arrival times differing by up to two seconds can still be tagged as coincidences.

A diagram of the coincidence matching algorithm is shown in \figref{fig:tsmatch}. Events from both the head and the tail frontend are placed into a buffer which orders the events based on their trigger time as measured by the TSC FIFO. Whenever a new event is placed into the buffer, the trigger time difference between the earliest and the latest event in the buffer is calculated. If this difference is greater than some set value (the default setting is four seconds), then the entire queue is searched for coincidence matches with the earliest event. Here, a match is defined as any two events whose timestamps are within $10~\mu$s of each other. Regardless or whether or not a match is found, the earliest event is sent to a singles event processor which calculates all of the necessary singles parameters and sends the event on to the next stage of analysis (which is typically either histogramming, writing to disk, or both). After this, the event is removed from the buffer.  If a coincidence match \emph{is} found, the matching events are also sent to a coincidence event processor. Note that in the case of coincidences, only the earliest event is removed from the buffer. The other event will remain until it becomes the earliest event, at which time it will be analyzed as a singles event and then removed.

In practice, an \code{std::\allowbreak{}multiset} from the \cpp{} standard library~\cite{cppstdlib} is used as the event buffer. This container automatically maintains sorting between elements, which results in very efficient searches for coincidence matches. Furthermore, the automatic sorting naturally lends itself to checking the time difference between the earliest and the latest event in the buffer. It also allows for multiple coincidences to be stored and tagged. This is not necessary at present since the dead times render multiple coincidences (within a $10~\mu$s window) impossible. However, it allows for easy expansion of the algorithm should multiple coincidences ever become possible. The performance of the \code{std::\allowbreak{}multiset} was checked against a variety of other options, including  a \code{std::\allowbreak{}vector} and \code{std::\allowbreak{}deque} which are resorted after every insertion and an unordered hash container, \code{boost::\allowbreak{}unordered\_multiset}~\cite{boosttr1}. The sorted \code{std::\allowbreak{}deque} performed similarly to the \code{std::\allowbreak{}multiset} for small objects. However, for objects the size of a real event, the additional copy operations involved in the resorting reduce performance significantly. The performance of the \code{std::\allowbreak{}multiset} was slightly worse than the \code{boost::\allowbreak{}unordered\_multiset} in terms of searching for coincidence matches. However, the difficulties associated with a lack of ordered elements in the latter container, as well as its reliance on non-standard libraries, do not justify the small performance increase.

\subsection{Live Times}
\label{sec:livetime}

To correctly measure the yield of a reaction, it is necessary to correct the number of recorded events for the live time of the \ac{DAQ}. This can be done by determining the fraction of time, $L$, during which the acquisition is open to new triggers. The real number of events, $N$, is then equal to the number of recorded events, $n$ divided by $L$,
\begin{equation}
\label{eq:livetime}
	N = n / L.
\end{equation}
In conventional systems with non-paralyzable dead times, $L$ can be determined simply from the sums of recorded scaler counts. For example, if it is possible to count the number of presented and accepted triggers, then $L$ is simply given by
\begin{equation}
\label{eq:livetime_scaler}
	L = N_{\text{acq}} / N_{\text{pres}}.
\end{equation}

In the \ac{DRAGON} \ac{DAQ}, there are two free-running systems with independent singles live times. For each of the singles triggers, the live time corrections can be made by the method outlined above. However, for coincidences this is not possible since coincidence tagging is performed at the analysis stage, making it impossible to count the rate of presented coincidences in a scaler alone.
%
The lack of an available method for counting presented coincidences means that other methods must be employed to determine live time corrections. One option is to directly measure the busy time associated with each recorded event, that is, to record how long the \ac{DAQ} is blind to incoming triggers on an event-by-event basis.  This is part of the standard operating procedure for the \iothirtytwo{}, which calculates the total busy time for each event from \ac{TSC} measurements and stores it in the data stream. For systems with a non-paralyzable dead time response and reactions generated as a random Poisson process, the number of events lost due to dead time, $n_{\text{lost}}$, is given by
\begin{equation}
\label{eq:nlost}
	n_{\text{lost}} = \lambda \sum_{i=0}^{n} \tau_i = \lambda \tau,
\end{equation}
where $n$ is the total number of \emph{recorded} events; $\lambda$ is the rate of \emph{generated} events; $\tau_i$ is the busy time associated with a given event $i$; and $\tau$ is the sum of all busy times across a run. The number of generated events, $N$, over the total run time $T$ is then given by
\begin{eqnarray}
	N	&=&	n + n_{\text{lost}} 					\label{eq:ntot1}	\\ 
		&=&	n + \lambda \tau						\label{eq:ntot2}	\\
		&=&	n + \left( N / T \right) \tau			\label{eq:ntot3} \\
		&=& \frac{n}{1 - \tau / T}.				\label{eq:ntot4}
\end{eqnarray}
From \eqnref{eq:ntot4}, it is straightforward to calculate the number of generated events from the number of recorded events, the measured dead times, and the total run time. Alternatively, it is possible to define the live time fraction as
\begin{equation}
\label{eq:livetime1}
	L  = 1 - \tau / T
\end{equation}
and then to use \eqnref{eq:livetime} to calculate $N$.

For a singles analysis, $\tau$ can simply be calculated as the sum of the individual $\tau_i$ over all events. For coincidences, however,  more care is required to calculate $\tau$ correctly. There are three classes of possibilities regarding the loss of coincidence events due to dead time:
\begin{enumerate}
	\item The event arrives when neither the head nor the tail is busy: the event will be recorded and tagged as a coincidence. \label{enum:coinc1}
	\item The event arrives when either the head or the tail is busy, but not both: half of the event will be recorded and tagged as singles.\label{enum:coinc2}
	\item The event arrives when both the head and the tail are busy: the event will not be recorded at all.\label{enum:coinc3}
\end{enumerate}
In terms of correcting recorded coincidence events for dead time losses, both cases \eqref{enum:coinc2} and \eqref{enum:coinc3} should count as a loss. Thus $\tau$ should be the total time during which the \emph{head or the tail} is busy (note that this is a true logical \logical{or} as opposed to the exclusive \logical{or} of case \eqref{enum:coinc2}).  To calculate $\tau$ for coincidence events, we employ an algorithm which stores the ``start'' and ``stop'' times of all busy periods from both \ac{DAQ}{}s, sorted by their start times. The algorithm then iterates through the list, identifies any cases of overlapping head and tail busy periods, and calculates the sum of busy times with the overlaps removed.

\subsubsection{Non-Poisson Events}
\label{sec:nonpoisson}

\newcommand{\isumfrac}{
	\frac{
			\sum_{i=0}^{n} \int_{0}^{\tau_i} R\of{t} \der t
		}{
			\int_{0}^{T} R\of{t} \der t
		}
}

The live time analysis presented in Eqns.~\eqref{eq:ntot1}--\eqref{eq:ntot4} is only valid when the rate of generated events is a Poisson process. This is usually the case when studying nuclear reactions such as radiative capture since the underlying physics adhere to Poisson statistics. However, in beam-based experiments such as those at \ac{DRAGON}, the reaction rate is governed by the underlying physics of the reaction, the rate of the incoming beam, and the target density. In cases where the beam rate (or target density) is fluctuating, the rate of occurrence of reactions becomes non-Poisson. Instead, the rate becomes an inhomogenous Poisson process, that is, one where the rate is time dependent. The expected number of events in a interval $[0,~ \tau]$ is then given by
\begin{equation}
	\label{eq:inhomogenous_poisson}
	\int_{0}^{\tau} \lambda \left( t \right) \der t.
\end{equation}
In an experiment, the time rate of reactions (assuming constant target density) is determined by the yield per incoming beam particle, $Y=N/N_b$, which is a constant, and incoming beam rate as a function of time, $R\of{t}$:
\begin{equation}
\label{eq:tdrate}
	\lambda\of{t} = \frac{N}{N_b} R\of{t}.
\end{equation}
The number of true events, $N = n + \nlost,$ is then
\begin{eqnarray}
	N	 	&=& n + \sum_{i=0}^{n} \int_{0}^{\tau_i} \frac{N}{N_b} R\of{t} \der t \\
			&=& n + N~ \isumfrac,
\end{eqnarray}
or solving explicitly for $N$:
\begin{equation}
	N = n \left( 1 - \isumfrac	\right)^{-1} .
\end{equation}
From here, we can define a live time fraction analogous to that of \eqnref{eq:livetime_scaler}:
\begin{equation}
\label{eq:livetimefull}
	L = 1 - \frac{\sum_{i=0}^{n} \int_{0}^{\tau_i} R\of{t} \der t }{\int_{0}^{T} R\of{t} \der t}.
\end{equation}
Note that if the beam rate is a constant with respect to time, $R\of{t} \equiv R$, we recover the definition of $L$ given in \eqnref{eq:livetime1}:
\begin{eqnarray}
	\int_0^{\tau} R\of{t} \der t &	= &	\int_0^{\tau} R \der t = R \tau 		\\
	\Rightarrow L 	& 	= & 1 - \frac{\sum_{i=0}^{n} R \tau_i}{R T}	\\
					&   = & 1 - \frac{\sum_{i=0}^{n}   \tau_i}{T}  	\\
					&   = & 1 - \tau / T.
\end{eqnarray}

In \ac{DRAGON} experiments, the beam rate is monitored continuously by measuring the rate of elastically scattered target nuclei with \ac{IIS} detectors (c.f. \secref{subsec:dragon}). Thus it is possible to construct $R\of{t}$ from these measurements and use the full form of \eqnref{eq:livetimefull} for live time corrections. In \secref{sec:livetime_analysis}, we discuss the effect of including this full live time calculation in the analysis of the \rxnfull{20}{Ne}{p}{\gamma}{21}{Na} data reported in  \citeref{PhysRevC.88.038801}. We show that the change in the live time after accounting for beam fluctuations is small even in the presence of substantial rate changes. However, as a general rule, the sensitivity of the final result of a measurement to higher-order live time effects will be different for each experiment. Hence the appropriate live time analysis must be considered on a case-by-case basis.

\section{DAQ Commissioning Experiment}
\label{sec:commissioning}

\begin{figure*}
\centering
\includegraphics{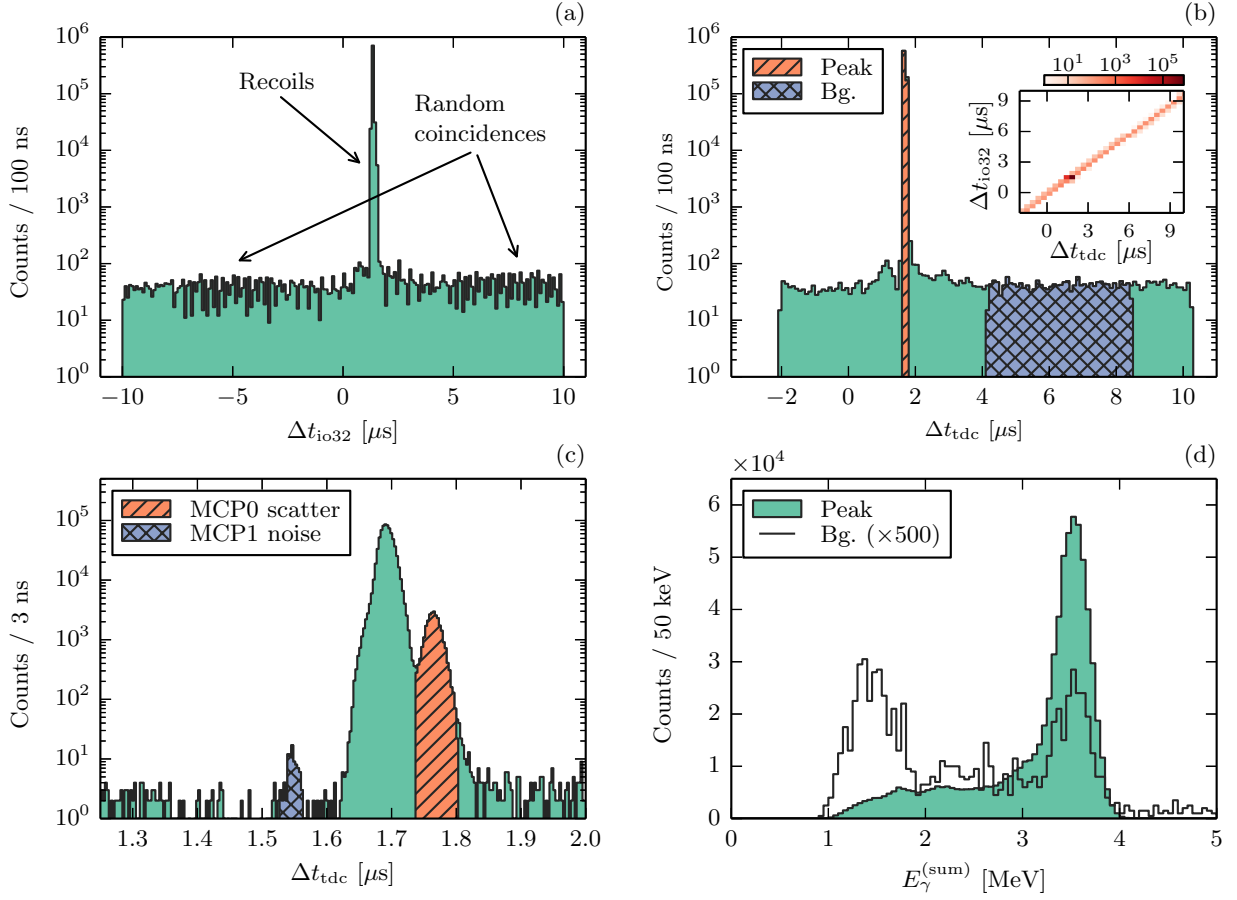}
\caption{
	\panel{a}: difference in trigger times (tail minus head) as measured by the \iothirtytwo{} \ac{TSC}.
	\panel{b}: difference in trigger times as measured by the head \ac{TDC}, with the hatched regions denoting cuts applied to the spectra in \panel{d}. The inset shows the correlation between the \iothirtytwo{} and \ac{TDC} trigger time difference measurements.
	\panel{c}: close up of the recoil peak region of the \ac{TOF} spectrum. The structure of the peaks and the meaning of the hatched regions are discussed in the text.
	\panel{d}: sum of coincidence $\gamma$-ray energies measured in the \ac{BGO} array. The shaded histogram consists of events from the recoil peak in separator \ac{TOF} (diagonal-hatched region in \panel{b}), while the unshaded histogram is composed of an arbitrary background region (cross-hatched region in \panel{b}).
}
\label{fig:coinc}       
\end{figure*}

The new \ac{DRAGON} \ac{DAQ} was commissioned by measuring the strength of the $\ecm = 1113$ keV resonance in the \rxnfull{20}{Ne}{p}{\gamma}{21}{Na} radiative proton capture reaction. This reaction was also used in the original \ac{DRAGON} commissioning experiment~\cite{Engel2005491, EngelThesis}. Since the separator hardware has not changed appreciably since its inception, revisiting this reaction provides a reliable means to check for any inconsistencies that might be introduced by the new \ac{DAQ}. This resonance also serves as an important calibration point for measurements of direct radiative capture in \rxnfull{20}{Ne}{p}{\gamma}{21}{Na} at lower energies. As the starting point of the NeNa cycle, these are important for the nucleosynthesis of intermediate mass elements in ONe classical novae and the production of sodium in yellow supergiants~\cite{Rolfs1975460, 1999ApJ...520..347J, 1991ApJ...379..729P, Bloch1969129, Iliadis201031}.

A heavy-ion singles analysis of the \ac{DAQ} commissioning experiment has already been reported in \citeref{PhysRevC.88.038801}, with the results proving consistent with the original \ac{DRAGON} commissioning. It was also shown that the commonly accepted value of the $\ecm = 1113$ keV resonance strength was incorrectly derived from a 1960 measurement~\cite{ThomasAndTanner} that was reported in the laboratory frame of reference and later misinterpreted as being in the center-of-mass frame. As shown in \citeref{PhysRevC.88.038801}, recalculating the resonance strength of \citeref{ThomasAndTanner} in the center-of-mass frame brings it into agreement with other published measurements~\cite{Engel2005491, PhysRevC.15.579}. As a result, it was recommended that the accepted value be lowered  to account for this new information.

Since we have already reported the singles analysis in \citeref{PhysRevC.88.038801}, here we focus on the coincidence aspects of the data. A summary of relevant coincidence parameters is presented in \figref{fig:coinc}.  \panel{a} shows the difference in trigger times for the head and tail \ac{DAQ} systems, as measured by their respective \iothirtytwo{} \ac{TSC}{}s. As indicated in the figure, the peak around \meas{1.5}{$\mu$s} consists of true recoil events. This sits on top of a flat random background resulting from accidental coincidences between a heavy ion and an uncorrelated $\gamma$ ray.  \panel{b} shows the difference in trigger times as measured by the head \ac{TDC}, which shows the same structure as the \iothirtytwo{} measurements. The inset shows the correlation between the \iothirtytwo{} and \ac{TDC} time difference measurements. As expected, they show a near $1$:$1$ correlation, with a slight offset due to differing signal propagation delays.

\panel{c} shows a zoomed-in view of the recoil peak in the separator \ac{TOF}, which reveals additional structure. The main peak around \meas{1.7}{$\mu$s} is made up of normal recoil events which are transmitted through both \ac{MCP}{}s to the \ac{DSSSD}. The small cross-hatched peak to the left of the main one consists of events in which a valid \ac{MCP}{}0 signal is coincident with noise in \ac{MCP}{}1. This is evidenced by looking at the relative timing of \ac{MCP}{}0 and \ac{MCP}{}1, which shows a random distribution of times for the \ac{MCP}{}1 trigger relative to \ac{MCP}{}0. The diagonal-hatched peak to the right of the main one likely consists of events which scatter in the carbon foil of \ac{MCP}{}0 and are transmitted to \ac{MCP}{}1 with a reduced velocity compared to unscattered recoils. These events have the same \ac{TOF} from the target to \ac{MCP}{}0 as events in the main peak, but their \ac{TOF} from \ac{MCP}{}0 to \ac{MCP}{}{1} is around \meas{60}{ns} longer, and it has a significantly broader distribution compared to normal recoils. This is indicative of events which originated as normal recoils at the target and then changed velocity due to some reaction process in \ac{MCP}{}0. For events that trigger off the \ac{MCP}{}s (as opposed to the \ac{DSSSD}), the \ac{MCP}{}1 signal defines the trigger, so as a result the separator \ac{TOF} is taken relative to \ac{MCP}{}1. This means that any delay in \ac{MCP}{}1 timing due to recoils changing velocity in \ac{MCP}{}0 will manifest as a delay in the separator \ac{TOF} by the same amount, which is the case for events in the cross-hatched peak. Furthermore, the cross-hatched events do not come with a valid signal in the \ac{DSSSD} as would be expected for recoils which scatter and change their trajectory to one outside the \ac{DSSSD} acceptance. 

\panel{d} in \figref{fig:coinc} shows the \ac{BGO} $\gamma$-ray energy sum for events in the recoil peak (shaded histogram) superimposed with events from an arbitrary background region outside the recoil peak (unshaded histogram). As expected, the recoil $\gamma$ rays are almost all concentrated in a strong peak at the \meas{3.5}{MeV} decay energy of the state populated in the reaction. The background histogram, on the other hand, has a significant enhancement near threshold resulting from room background $\gamma$-rays.

\subsection{Resonance Strength Calculation}
\label{sec:wgcoinc}

\newcommand{\numBeam}		{\ee{\left( 2.296 \pm 0.032 \right)}{15}}
\newcommand{\numRecoil}	{\ee{\left( 5.923 \pm 0.062 \right)}{5}}
\newcommand{\effTrans}		{94 \pm 3 \%}
\newcommand{\effLive}		{91.486 \pm 0.002 \%}
\newcommand{\effBgo}		{55.9 \pm 10.0  \%}
\newcommand{\effTot}		{21.1 \pm 3.9 \%}
\newcommand{\yieldC}		{\ee{\left(1.221 \pm 0.221 \right)}{-9}}
\newcommand{\wgC}			{0.969 \pm 0.210}

\newcommand{\yieldS}		{\ee{\left(1.225 \pm 0.051 \right)}{-9}}
\newcommand{\wgS}			{0.972 \pm 0.119}

\begin{figure}
\centering
\includegraphics{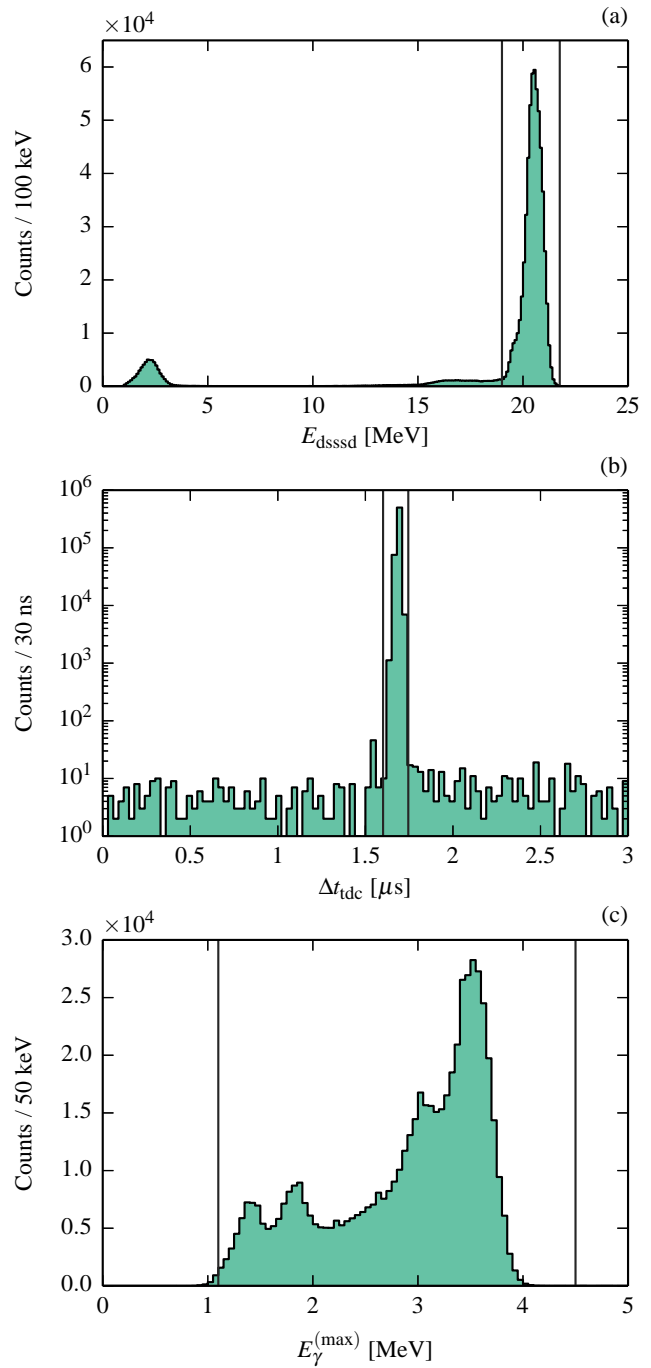}
\caption{
Summary of cuts used in the coincidence resonance strength analysis.
\panel{a}: \ac{DSSSD} energy cut; \panel{b}: separator \ac{TOF} cut on the prompt recoil peak; and \panel{c}: cut on the energy deposited by the most energetic $\gamma$-ray. In each panel, the cut limits are indicated by vertical lines. The final recoil cut is an \logical{and} of all three conditions. All histograms consist of coincidence events only, and those in Panels (b) and (c) consist only of events which pass the cuts shown in the panels above them.
}
\label{fig:recoilgates}
\end{figure}

To verify that the timestamp-based coincidence matching is working as intended, we have performed a full coincidence analysis of the $\ecm = 1113$ keV resonance strength in \rxnfull{20}{Ne}{p}{\gamma}{21}{Na}, using data taken during the \ac{DAQ} commissioning experiment. The details of the experiment and the resonance strength calculation, including the employed stopping power, are identical to \citeref{PhysRevC.88.038801}. However, the recoil event selection and overall efficiency are different in the present analysis. A summary of the recoil event selection is shown in \figref{fig:recoilgates}. The final recoil cut is an \logical{and} of the \ac{DSSSD} energy cut used in \citeref{PhysRevC.88.038801}, a cut on the recoil peak in separator \ac{TOF}, and a cut on the energy deposited by the most energetic $\gamma$-ray.

\tableref{tab:yield} shows a summary of the detection efficiency, yield, and resonance strength in the coincidence analysis. The detection efficiency differs from that of \citeref{PhysRevC.88.038801} in two ways: the live time is different as a result of the coincidence trigger requirement, and the $\gamma$-ray detection efficiency becomes part of the total efficiency product.  The live time was calculated using \eqnref{eq:livetime1}, with the total dead time $\tau$ being the logical \logical{or} of dead times in the head and tail systems, as explained in \secref{sec:livetime}. The uncertainty on the live time is equal to the $20$ parts per million accuracy rating of the \iothirtytwo{} quartz crystal, i.e. a \emph{relative} uncertainty of $0.002 \%.$  The \ac{BGO} efficiency was calculated from a \GEANT{3} simulation~\cite{darioThesis}, with the branching ratios for the decay of the \meas{3.54}{MeV} state in \nuc{21}{Na} taken from \citeref{Firestone2004269}. The simulated events were analyzed with the same energy cut as the data:
\begin{equation*}
	1.1~\text{MeV} < E_{\gamma}^{\mathrm{(max)}} < 4.5~\text{MeV},
\end{equation*}
where $E_{\gamma}^{\mathrm{(max)}}$ is the energy deposited by the most energetic $\gamma$ ray. This accounts for any effect of hardware thresholds since the lower limit of \meas{1.1}{MeV} is beyond the range of the threshold function. The uncertainty on the \ac{BGO} efficiency calculation was estimated at $10\%$ as explained in \citeref{darioThesis}.

We have also assigned an uncertainty of $3\%$ to the gas target transmission. This was estimated from the standard deviation of Faraday cup readings taken upstream and downstream of the gas target. Each set of readings sampled the beam current approximately every \meas{0.2}{s} over the course of \meas{30}{s.}
This transmission uncertainty was not included in the singles analysis of \citeref{PhysRevC.88.038801}. In \tableref{tab:yield} we also report the updated singles yield and resonance strength when including the $3\%$ gas target transmission uncertainty in the calculation. The overall effect is minor, appearing only in the last quoted digit of the uncertainty on the resonance strength.

As indicated in \tableref{tab:yield}, the present yield and resonance strength are in very good agreement with the singles values, as well as the measurements of Refs.~\cite{Engel2005491, ThomasAndTanner, PhysRevC.15.579} (with the appropriate center-of-mass corrections made to \citeref{ThomasAndTanner}). Note that the increased uncertainty on the coincidence yield and resonance strength as compared to their singles counterparts is a consequence of the $10\%$ uncertainty attached to the \ac{BGO} efficiency.

\newcommand{\sdiff}[1]{\textbf{#1}}
\newcommand{\sdifftype}{bold face}
\begin{table}
\caption{
Summary of the coincidence yield calculation. Quantities which are different from or supplementary to those in \citeref{PhysRevC.88.038801} are labeled in \sdifftype{}. The table also includes the singles yield and resonance strength, with the uncertainties on these quantities updated relative \citeref{PhysRevC.88.038801}. They now include propagation of the $3\%$ uncertainty on the gas target transmission.
}
\label{tab:yield}

\begin{tabular}{ll}
\noalign{\smallskip}\hline
\hline\noalign{\smallskip}
Quantity & Value  \\
\noalign{\smallskip}\hline\noalign{\smallskip}
\ac{DSSSD} detection efficiency					\cite{Wrede2003619}				& $97.0 \pm 0.7\%$	    \\
Ne$^{\left(9+\right)}$ charge state fraction		\cite{Engel2005491}				& $59 \pm 1\%$ 		     \\
DRAGON transmission 								\cite{EngelThesis}				& $99.9^{+0.1}_{-0.2}\%$ 	    \\
\ac{MCP} transmission								\cite{Vockenhuber2009372}		& $76.9 \pm 0.6\%$	   \\
\sdiff{Gas target transmission}						& $\effTrans$	 \\
\sdiff{Coincidence live time} 							& $\effLive$	 \\
\sdiff{BGO detection efficiency}						& $\effBgo$ 	 \\
\sdiff{Total coincidence efficiency} 					& $\effTot$ 	 \\
\noalign{\smallskip}\hline\noalign{\smallskip}
Integrated beam current 								& $\numBeam$ 		\\
\sdiff{Detected recoils}								& $\numRecoil$ 		\\
\sdiff{Yield}											& $\yieldC$ 		\\
\sdiff{Resonance strength} 		 					& $\wgC$ eV 	\\
\noalign{\smallskip}\hline\noalign{\smallskip}
\sdiff{Singles yield}	 								& $\yieldS$ \\
\sdiff{Singles resonance strength} 					& $\wgS$ eV \\

\noalign{\smallskip}\hline
\noalign{\smallskip}\hline
\end{tabular}
\end{table}

\subsection{Live Time Analysis}
\label{sec:livetime_analysis}
\newcommand{\binOne}{\Delta}
\newcommand{\binTwo}{\Delta_{30}}

The rate of the incoming \nuc{20}{Ne} beam varied significantly throughout the course of the \ac{DAQ} commissioning experiment. This is because the beam was extracted from the ISAC offline microwave ion source \cite{jayamanna:02C711}, which requires passing the beam through a stripper foil to reach a charge state suitable for acceleration. Degradation of stripper foils resulted in steady decreases in beam intensity on the time scale of a few hours, after which the ISAC operators would replace the foil and return the beam intensity to its initial state. As a result, the present data set provides an ideal case to examine the effect of varying beam rates on the live time calculations.

To examine the effect of varying beam rates on the live time, we calculated heavy-ion \emph{singles} live times on a run-by-run\footnote{Each run represents \meas{\sim 1}{hour} of data taking bookended by Faraday cup readings.} basis using three different methods. The first method (``Poisson'') calculates the live time from the total sum of busy times divided by the total run time, as represented by \eqnref{eq:livetime1}. This method was employed both in the present coincidence resonance strength calculation (\secref{sec:wgcoinc}) and the singles result reported in \citeref{PhysRevC.88.038801}. The second method (``scaler'') uses the ratio of acquired to presented triggers measured by the \iothirtytwo{} scalers, viz.\ \eqnref{eq:livetime_scaler}.  The third method (``non-Poisson'') involves treating the trigger rate as an inhomogeneous Poisson process as outlined in \secref{sec:nonpoisson}. For this method, we treat the incoming beam rate, $R\of{t},$ as being proportional to the the \ac{IIS} trigger rate, $R_{\mathrm{iis}}\of{t}.$  The proportionality constant cancels out in \eqnref{eq:livetimefull}, so we can set $R\of{t} \equiv R_\mathrm{iis}\of{t}.$
To evaluate \eqnref{eq:livetimefull} from the measured \ac{IIS} rates, we divide each run into \meas{30}{second} periods. For each period $j$ we treat the rate as being a constant $R_j$ equal to the average \ac{IIS} rate over the period. To evaluate the integrals over $R\of{t} \der t,$ we use a simple rectangle method with one second bin sizes.
Expressed mathematically, our approximation to \eqnref{eq:livetimefull} is
\begin{equation}
\label{eq:livetimefull_numerical}
	L \simeq 1 - 
	\frac{
			\sum_{j=0}^{n_j} \left. R_{j}  \sum_{i:~ \tau_i \in S_j}\tau_{i} \right.
		 }{
		 	\sum_{i=0}^{n_i} R_i \binOne
		 },
\end{equation}
where the index $i$ corresponds to the one second bins used for integral evaluation; $\binOne$ is equal to one second; $n_i$ is the number of one-second divisions per run, i.e. $T/\binOne$; $R_i$ is the measured \ac{IIS} rate during each bin $i$; and $\tau_i$ is the sum of measured busy times within the bin $i$. The index $j$ is over the \meas{30}{second} periods during which we treat the \ac{IIS} rate as constant, and $n_j$ is the number of $30$ second periods per run. $S_j$ is the time interval corresponding to a period $j$, i.e. the range $t_j \leq t \leq t_j + 30~${}s where $t_j$ deontes the start of the period $j$. Finally, $R_j$ is the average \ac{IIS} rate over a division $j$:
\begin{equation}
\label{eq:r30sec}
	R_{j}  \simeq
			\frac{
					\sum_{k :~ t \in S_j} R_k \binOne
				}{
					\sum_{k :~ t \in S_j} \binOne
				}
					 .
\end{equation}

\begin{figure}
\centering
\includegraphics{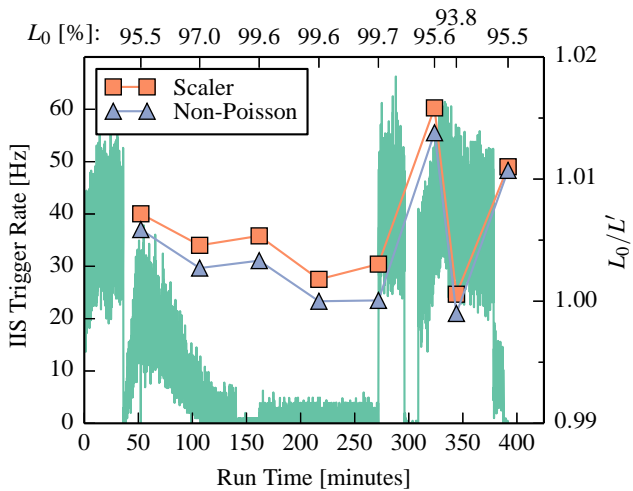}
\caption{
	Effect of beam fluctuations on the heavy-ion singles live time calculations. The solid histogram shows the \ac{IIS} trigger rate as a function of time. The filled squares and triangles denote the ratio of live times $L_0/L^{\prime},$ where $L_0$ is the live time calculated using \eqnref{eq:livetime1} and $L^{\prime}$ is the live time calculated using an alternative method. The alternative methods are the ``scaler'' method (squares) using  \eqnref{eq:livetime_scaler} and the ``non-Poisson'' method (triangles) using \eqnref{eq:livetimefull}.  The horizontal positions of the markers denote the end times of runs throughout the experiment. The run-by-run live times calculated using the Poisson method are also displayed textually across the top of the figure.
}
\label{fig:livetime}
\end{figure}

The results of the live time analysis are shown in \figref{fig:livetime}, along with the \ac{IIS} trigger rate as a function of time. The figure presets the scaler and non-Poisson live times as ratios to the Poisson live time, $L_0 / L^\prime,$ where $L_0$ is the Poisson live time and $L^\prime$ is either the scaler or non-Poisson live time. This represents the fractional change in the yield that would result from using either the scaler or non-Poisson  live time instead of the Poisson. The scaler and non-Poission live times agree well with each other. For the runs with the most significant rate fluctuations, they trend towards being slightly lower than the Poisson live time (resulting in a higher $L_0 / L^\prime$ ratio) but still differ by no more than $1.5\%.$ We have also calculated  live times across the entire experiment, $L_{\mathrm{full}},$ by taking the weighted average of the inverse of the run-by-run live times with the weights being the number of recoils detected:

\begin{equation}
\label{eq:overall_livetime}
	L_{\mathrm{full}}
		 = \left( \frac{
				\sum_{i} n_{r,i} / L_i
			} {
				\sum_{i} n_{r,i}
			} \right)^{-1}.
\end{equation}
Calculated this way, correcting the total sum of detected recoils using $L_{\mathrm{full}}$ is the mathematical equivalent of making live time corrections run-by-run, i.e.
\begin{equation}
\label{eq:lfull}
	\sum_i \frac{n_{r,i}}{L_{i}} = \frac{\sum_i n_{r,i}}{L_{\mathrm{full}}}.
\end{equation}
The respective $L_{\mathrm{full}}$ values for the Poisson, scaler, and non-Poisson methods are $95.6\%,$ $94.8\%$, and $94.9\%$. These translate to fractional yield changes $(Y^\prime / Y_0 - 1)$ of $0.86\%$ and $0.71\%$ for the scaler and non-Poisson methods, respectively. We have also calculated the non-Poisson live time for the coincidence measurement presented in \secref{sec:wgcoinc}, arriving at an overall live time of $90.7\%$, which translates to a fractional yield change of $0.91\%.$
Such changes are small compared to the overall error budget. Since the present analysis represents a particularly extreme case of beam rate fluctuations, this can be taken as an indication that the final result of a \ac{DRAGON} experiment is not likely to be sensitive to the particular method of live time calculation.

\section{Conclusions}
\label{sec:conclusions}

In conclusion, we have developed a new \ac{DAQ} for the \ac{DRAGON} recoil mass separator at \triumf{}. The new \ac{DAQ} consists of two free-running acquisition systems with completely independent triggering and readout, one for the $\gamma$-ray detectors surrounding the target and the other for heavy-ion detectors at the end of the separator.  Events are recorded with timestamps from a local \meas{20}{MHz} clock, with the clock frequencies and zero-point offsets synchronized between the two systems. Comparison of timestamp values allows coincidence events to be identified in the first stage of data analysis, and we have implemented and successfully employed a coincidence-matching algorithm that is suitable for both online and offline analysis.  

The new \ac{DRAGON} \ac{DAQ} was commissioned by measuring the strength of the \meas{\ecm = 1113}{keV} resonance in the \rxnfull{20}{Ne}{p}{\gamma}{21}{Na} radiative capture reaction. The experiment ran successfully, and the measured coincidence resonance strength, \meas{\wg = \wgC}{eV}, is in good agreement with our previous singles result~\cite{PhysRevC.88.038801}, as well as earlier publications~\cite{Engel2005491, ThomasAndTanner, PhysRevC.15.579}.  All activities to date indicate that the \ac{DAQ} upgrade is successful and that the new system can be used in future \ac{DRAGON} experiments.

\section{Acknowledgements}
\label{sec:acknowledgement}

We are grateful to the ISAC operations and offline ion source groups for delivery of a high quality \nuc{20}{Ne} beam during the \ac{DAQ} commissioning experiment. We also thank P. Amadruz for his guidance and efforts in developing the new \ac{DAQ} system. This work was supported in part by the National Research Council and National Sciences and Engineering Research Council of Canada.

 \bibliographystyle{apsrev4-1}
 \bibliography{./daq_references}{}

\end{document}